\documentstyle[epsfig,12pt]{article}
\newcommand{\td}{\textstyle}

\newcommand{\tf}{\over\textstyle}

\begin{document}
\author{ Andr\'e Roug\'e\thanks{Andre.Rouge@in2p3.fr}\\
 LPNHE Ecole Polytechnique-IN2P3/CNRS\\
 F-91128 Palaiseau Cedex}
\date{August 1997}
\title{Isospin constraints on the $\tau\to K\bar Kn\pi\nu$
decay mode}
\maketitle
\begin{abstract}
The construction of the complete isospin relations and inequalities between the
possible charge configurations of a $\tau\to K\bar Kn\pi\nu$
decay mode is presented. Detailed applications to  the cases of two and three pions
are given.
\end{abstract}
\mbox{~}\\[-11cm]
\begin{flushright}
X-LPNHE 97/08
\end{flushright}
\mbox{~}\\[8.5cm]
\section{Introduction}

The isospin constraints on the $\tau$ hadronic decay modes are known
for the $n\pi$ and $Kn\pi$  modes\cite{gil,ar}.
For the   $K\bar Kn\pi$ final states, only the simplest decay mode $K\bar K\pi$
has been studied.
Using the formalism of symmetry classes introduced by Pais\cite{pais}, we generalize 
the relations for an arbitrary value of $n$ and give a geometrical representation of the
constraints.
\section{The general method}
The $ K\bar K n\pi$ system produced by a $\tau$ decay has isospin 1; the
possible values of the $K\bar K$ isospin $I_{K\bar K}$ are 0 and 1
and the isospin $I_{n\pi}$ of the $n$ pion system  is 1 for
 $I_{K\bar K}=0 $ and 0, 1 or 2 for  $I_{K\bar K}=1 $.

 Since there
is no second-class current in the Standard Model, interferences
between amplitudes with $I_{K\bar K}=0$ and  $I_{K\bar K}=1$
vanish in the partial widths \cite{ar}. Therefore we have  the relation
\begin{equation}
\Gamma_{K^0\bar K^0(n\pi)^-}=\Gamma_{K^+K^-(n\pi)^-},
\end{equation}
which is true for each charge configuration of the $n\pi$ system, and, since $\Gamma_{K_SK_S(n\pi)^-}=\Gamma_{K_LK_L(n\pi)^-}$,
\begin{equation}
\Gamma_{K^+K^-(n\pi)^-}=\Gamma_{K_S K_L(n\pi)^-}+2\Gamma_{K_S K_S(n\pi)^-},
\end{equation}
using the most easily observable states.

The amplitudes are classified by the values of $I_{K\bar K}$ and $I_{n\pi}$. 
To complete the classification, we use the isospin symmetry class\cite{pais} of the $n\pi$
system i.e. the  representation of
the permutation group ${\cal S}_n$ to which belongs the state. It is
characterized by the lengths of the three rows of its Young diagram $(n_1n_2n_3)$.
Due to the Pauli principle, the momentum and isospin states have the same symmetry. 
Thus integration over the momenta kills the interferences between amplitudes in different 
classes and there is no contribution from them in the partial widths.
 
Since $I_{n\pi}=0$ and $I_{n\pi}=1$ amplitudes belong to different symmetry classes\cite{pais},their interferences vanish.
 The presence of  $I_{n\pi}=2$ amplitudes makes the problem more intricate since they
share symmetry properties with some  $I_{n\pi}=1$ or  $I_{n\pi}=0$ amplitudes \cite{pais2,pk}.
For instance, in the case   $n=2$ the symmetry class $(200)$ is shared by  $I_{n\pi}=0$
and  $I_{n\pi}=2$; in the case $n=3$,  the symmetry class $(210)$ is shared by  
$I_{n\pi}=1$ and  $I_{n\pi}=2$.
Therefore the  allowed domains in the
space of the charge configuration  fractions 
($f_{cc}=\Gamma_{cc}/\Gamma_{ K\bar K n\pi}$)  must be determined separately for
each symmetry class and $I_{K\bar K}$, taking interferences into account when necessary.

 The complete
allowed domain is the convex hull of the  sub-domains corresponding to the different  $I_{K\bar K}$ and
symmetry classes and its projections are the convex hulls of their
projections.

For $n\leq 5$, which is always true in a $\tau$ decay, the isospin values
 and the symmetry class characterize unambiguously the  
amplitude properties \cite{pais2,pk}, therefore there is, at most, one interference term per class.
The sub-domain, for such a symmetry class associated with two different $I_{n\pi}$ values,
 is then a two-dimensional one since the partial widths $\Gamma_{cc}$ are linear functions of
three quantities: the sums of squared amplitudes for the two values of   $I_{n\pi}$ and the interference
term.
Its   boundary is determined by the Schwarz's inequality \cite{lm}.
This boundary is an ellipse; it  can be parametrized by writing  
the sums of squared amplitudes for the
two values of $I_{n\pi}$ as $\,\rho[1\pm\cos\theta]/2\,$ and the largest interference term allowed by the
Schwarz's inequality as $\,k\rho\sin\theta\,$, where the coefficient $k$ depends
 on the coupling coefficients.
The most general domain is hence the convex hull of a set of points corresponding to the symmetry classes
without $I_{n\pi}=2$ and a set of ellipses. 

The cases $n=2$ and $n=3$ are presented in detail in the following sections.  They  both  have the property that
only one symmetry class is associated with two isospin values.
 Higher values of $n$ are not expected, for some time,  to be of experimental interest. 

\section{The decay~\boldmath $\tau\to K\bar K2\pi\nu$}

The possible states that can be observed for a $ \tau\to\nu K\bar K\pi\pi$ decay  are the following:
\begin{center}
$K_S K_S\pi^0\pi^-$\\$K_S K_L\pi^0\pi^-$\\$K_L K_L\pi^0\pi^-$\\
$K^+ K^-\pi^0\pi^-$\\$K^+\bar K^0\pi^-\pi^-$\\$K^0 K^-\pi^+\pi^-$\\
$K^0 K^-\pi^0\pi^0$.
\end{center}
As mentioned before, not all the corresponding partial widths are  independent
and we can use the four fractions: $2f_{K^+ K^-\pi^0\pi^-}=f_{K^+ K^-\pi^0\pi^-}+
f_{K^0 \bar K^0\pi^0\pi^-}$, $f_{K^+\bar K^0\pi^-\pi^-}$, $f_{K^0 K^-\pi^+\pi^-}$
and $f_{K^0 K^-\pi^0\pi^0}$, whose sum is equal to 1, to describe the possible
charge configurations in a three-di\-men\-sio\-nal space. 
The ratio $K_SK_S/K_SK_L$ is a free parameter independent of the charge
 configuration fractions.

The partial widths for all the charge configurations can be expressed as functions
of the positive quantities $S_{[I_{K\bar K},I_{\pi\pi}]}$ which are the  sums of the squared absolute values of the
amplitudes with the given  values of the isospins and  the interference term $\cal I$ of the $I_{\pi\pi}=0$ and $I_{\pi\pi}=2$ amplitudes.
With only two pions the coefficients
are readily obtained from a  Clebsch-Gordan table and we get
\begin{eqnarray}\Gamma_{K^+ K^-\pi^0\pi^-}+\Gamma_{K^0 \bar K^0\pi^0\pi^-}&=&
2\Gamma_{K^+ K^-\pi^0\pi^-}= S_{[0,1]}+{\td 1\tf 2}S_{[1,1]}+{\td 3\tf 10}S_{[1,2]}\nonumber\\
\Gamma_{K^+\bar K^0\pi^-\pi^-}&=&{\td 6\tf 10}S_{[1,2]}\\
\Gamma_{K^0 K^-\pi^+\pi^-}&=&{\td 2\tf 3}S_{[1,0]}+{\td 1\tf 2}S_{[1,1]}+
{\td 1\tf 30}S_{[1,2]}+{\cal I}\nonumber\\
\Gamma_{K^0 K^-\pi^0\pi^0}&=&{\td 1\tf 3}S_{[1,0]}+{\td 2\tf 30}S_{[1,2]}-{\cal I}.\nonumber 
\end{eqnarray}
The partial width $\Gamma_{K\bar K\pi\pi}$ is the sum over the charge configurations:
\begin{equation}
\Gamma_{K\bar K\pi\pi}=\sum_{cc}\Gamma_{cc}=\sum_{sc}S_{sc}.
\end{equation}
The Schwarz's inequality
bounding the interference term $\cal I$ is
\begin{equation}
|{\cal I}|\leq {\td 2\tf 3\sqrt{5}}\sqrt{S_{[1,0]}S_{[1,2]}}.
\end{equation}

Since the partial width $\Gamma=\Gamma_{K\bar K\pi\pi}$ is independent of the interference,
the above equations are also true for the fractions $f_{cc}=\Gamma_{cc}/\Gamma$,
replacing the $S_{sc}$ by the weights $W_{sc}=S_{sc}/\Gamma$ and normalizing the interference term. 
 
For $[I_{K\bar K} ,I_{\pi\pi}]$ equal to  $[0,1]$ and  $[1,1]$, the sub-domains 
are merely points we will  refer to as $[0,1]$ and  $[1,1]$.

For the two interfering classes $I_{K\bar K}=1$, $I_{\pi\pi}=0~\hbox{or}~2$ ($ W_{[0,1]}=W_{[1,1]}=0$), 
the sub-domain  is the
two-dimensional domain in the plane $f_{K^+\bar K^0\pi^-\pi^-}=4f_{K^+ K^-\pi^0\pi^-}$
bounded by the ellipse of equation ${\cal I}=\pm{\cal I}_{max}$:
\begin{eqnarray} \lefteqn{{\td 3\tf 4}\,(
f_{K^0 K^-\pi^+\pi^-}-2 f_{K^0 K^-\pi^0\pi^0}+{\td 1\tf 6}
f_{K^+\bar K^0\pi^-\pi^-})^2=}\hspace{4cm}\\&&f_{K^+\bar K^0\pi^-\pi^-}\,(
f_{K^0 K^-\pi^+\pi^-}+f_{K^0 K^-\pi^0\pi^0}-{\td 1\tf 6} f_{K^+\bar K^0\pi^-\pi^-}).\nonumber
\end{eqnarray}
As explained before, the complete domain is the convex hull of this ellipse and
the two points $[0,1]$ and $[1,1]$.
Calling $I$ the point of the ellipse for which $f_{K^0K^-\pi^0\pi^0}=0$, the domain is made of the 
tetrahedron having the points I, $[0,1]$, $[1,1]$ and $[1,0]$ for vertices and of the two half-cones
whose bases are the halves of the ellipse delimited by the points $I$ and $[1,0]$ and whose vertices are the points 
$[1,1]$ and $[0,1]$ respectively.
\begin{figure}[ht]
\begin{center}
\mbox{~}\\[-1cm]
\begin{tabular}{cc}
\mbox{\epsfig{file=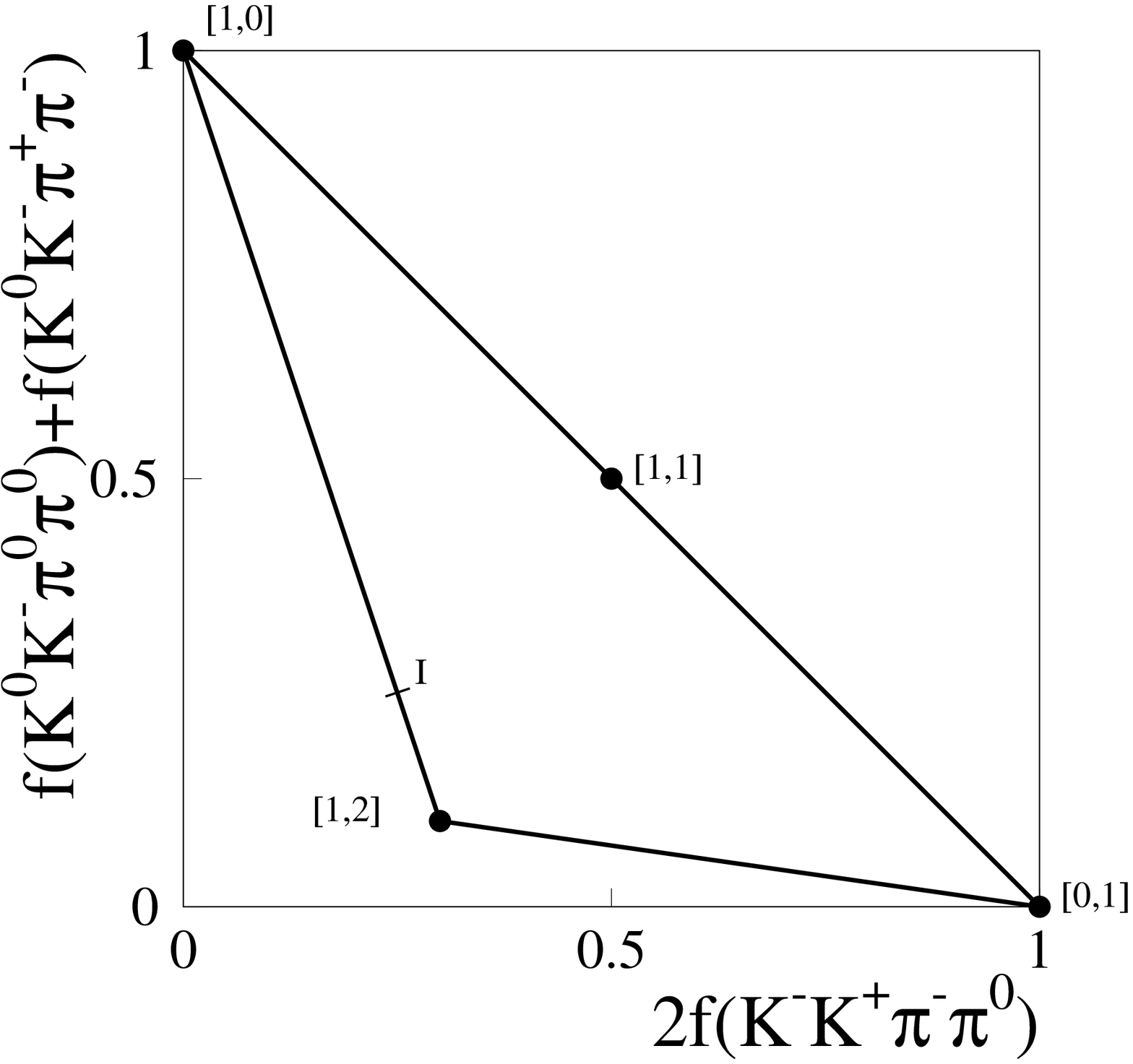,width=8cm}}&
\mbox{\epsfig{file=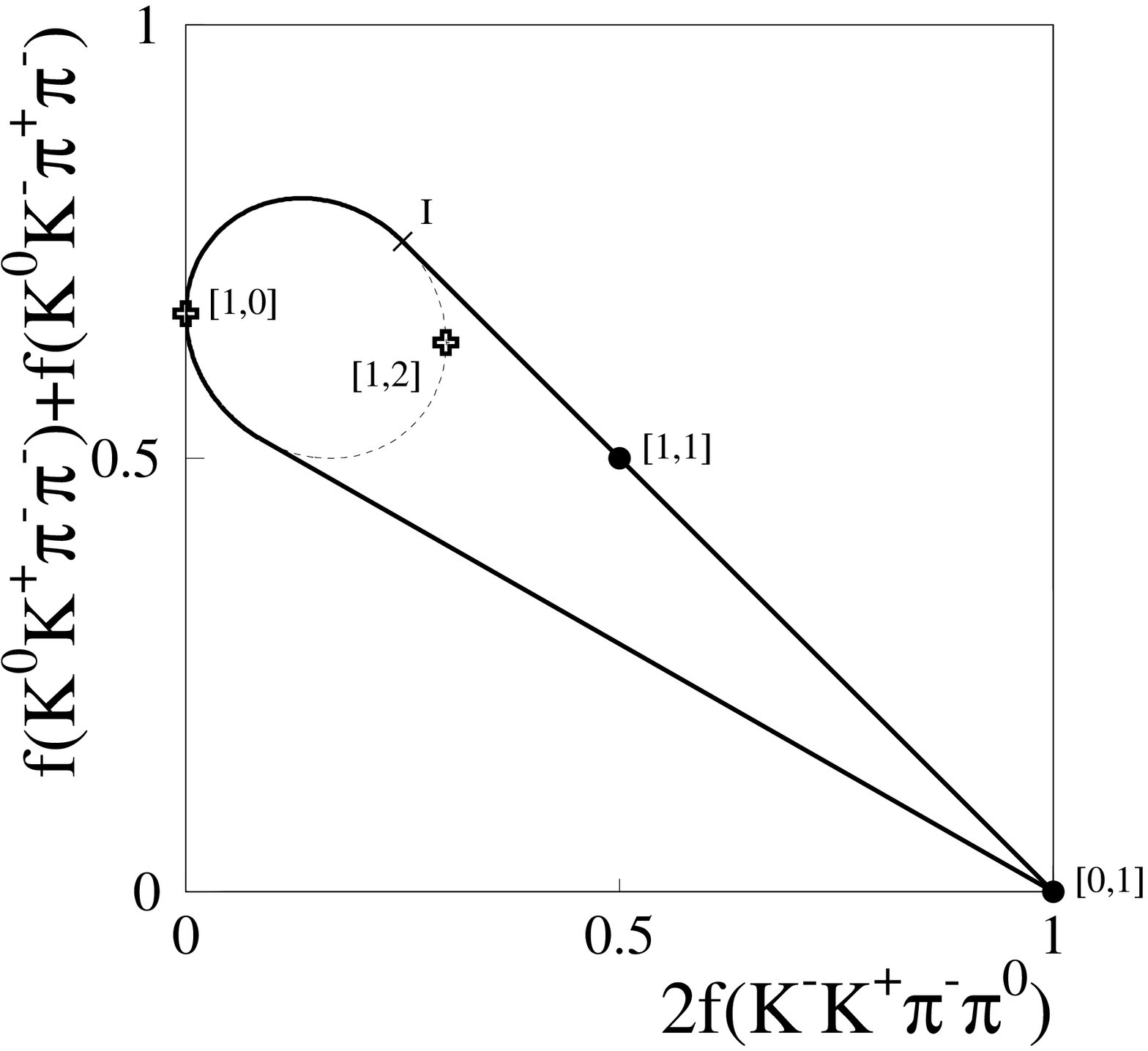,width=8cm}}
\end{tabular}
\end{center}
\caption{
Projections of the allowed domain on the planes $x/y$,
$x=2f_{K^+ K^-\pi^0\pi^-}$,\break  $y=f_{K^0 K^-\pi^+\pi^-}+f_{K^0 K^-\pi^0\pi^0}$ and
$x=2f_{K^+ K^-\pi^0\pi^-}$, $y=f_{K^+\bar K^0\pi^-\pi^-}+f_{K^0 K^-\pi^+\pi^-}$.
The classes of amplitudes are labelled  by the isospin values, $[I_{K\bar K},I_{2\pi}]$.
}
\end{figure}

For practical purposes it is useful to draw the projections of the domain on 
two-dimensional planes. The method is very simple: we first draw the projection
of the ellipse on the plane and then the tangents to the projected ellipse 
from the projections of the points $[0,1]$ and $[1,1]$.

A first simple example is the projection on the plane $x/y$, with $x=2f_{K^+ K^-\pi^0\pi^-}$
and $y=f_{K^0 K^-\pi^+\pi^-}+f_{K^0 K^-\pi^0\pi^0}$. Here the ellipse projection is a 
mere segment and the projected domain is the polygon whose vertices are the (projected)
points $[0,1]$, $[1,1]$, $[1,0]$ and $[1,2]$.

More interesting is the projection $x=2f_{K^+ K^-\pi^0\pi^-}$,
$y=f_{K^+\bar K^0\pi^-\pi^-}+f_{K^0 K^-\pi^+\pi^-}$, since the two final states
$K^+\bar K^0\pi^-\pi^-$ and $K^0 K^-\pi^+\pi^-$ have the same topology: one $K^0$ 
and three charged hadrons. The complement $1-x-y$ is the fraction $f_{K^0 K^-\pi^0\pi^0}$ 
of decays with two $\pi^0$'s.
The projected ellipse has vertical tangents at the points  $[1,0]$ and $[1,2]$.
It is also tangent to the line $x+y=1$ at the point $I$  ($x=1/4$), for which $W_{[1,0]}=W_{[1,2]}/5$
since   $f_{K^0 K^-\pi^0\pi^0}$ can be 0, because of the interference,  only when the two
contributions have the same modulus. The second tangent from $[0,1]$ touch the 
ellipse at the point of coordinates $x=2/23$ and  $y=12/23$. The allowed domain is shown on Fig.1.
The main constraint is the inequality
\begin{equation}
f_{K^0 K^-\pi^0\pi^0}\leq{\td 3\tf 4}(f_{K^+\bar K^0\pi^-\pi^-}+f_{K^0 K^-\pi^+\pi^-}),
\end{equation}
corresponding to the second tangent.
One can see from the plot that a large value of the ratio
$$(f_{K^+\bar K^0\pi^-\pi^-}+f_{K^0 K^-\pi^+\pi^-})/f_{K^+ K^-\pi^0\pi^-}$$
implies the dominance of $I_{K\bar K}=1 $ and a small value the dominance of
 $I_{K\bar K}=0 $. With one dominant isospin for the $K\bar K$ system,
the ratio $K_SK_S/K_SK_L$ measures the proportions of the two G-parities
i.e. the  contributions of axial and vector currents.
\section{The decay~\boldmath $\tau\to K\bar K 3\pi \nu$}
The final states for the decay $\tau\to K\bar K\pi\pi\pi\nu$ are:
\begin{center}
$K^0\bar K^0\pi^+\pi^-\pi^-$\\
$K^+ K^-\pi^+\pi^-\pi^-$\\
$K^0\bar K^0\pi^0\pi^0\pi^-$\\
$K^+ K^-\pi^0\pi^0\pi^-$\\
$K^0K^-\pi^+\pi^-\pi^0$\\
$K^0K^-\pi^0\pi^0\pi^0$\\
$K^+\bar K^0\pi^-\pi^-\pi^0$.
\end{center}

The relations between the $K^0\bar K^0$ and $K^+K^-$ final states are the same as in the 
\mbox{$\tau\to K\bar K\pi\pi\nu$} decay. Thus the charge configurations are described in a four-dimensional
space by the five  fractions:
$2f_{K^+ K^-\pi^+\pi^-\pi^-}=f_{K^+ K^-\pi^+\pi^-\pi^-}+f_{K^0\bar K^0\pi^+\pi^-\pi^-}$,
$2f_{K^+ K^-\pi^0\pi^0\pi^-}=f_{K^+ K^-\pi^0\pi^0\pi^-}+f_{K^0\bar K^0\pi^0\pi^0\pi^-}$,
$f_{K^0K^-\pi^+\pi^-\pi^0}$, $f_{K^0K^-\pi^0\pi^0\pi^0}$ and $f_{K^+\bar K^0\pi^-\pi^-\pi^0}$.

We shall label the amplitudes by the two isospin values and the symmetry class: $[I_{K\bar K},(n_1n_2n_3)^{I_{3\pi}}]$.
The relations between the partial widths for the charge configurations and the amplitudes
use both Clebsch-Gordan coefficients and the similar coefficients for the 
symmetry classes \cite{pais2,pk}. With the notations defined in the previous section, they can be written 
\begin{eqnarray}
2\Gamma_{K^+ K^-\pi^-\pi^0\pi^0}&=&\Gamma_{K^0\bar K^0\pi^-\pi^0\pi^0}+\Gamma_{K^+ K^-\pi^-\pi^0\pi^0}\nonumber\\
&=& {\td 1\tf 5}S_{[0,(300)^1]}+{\td 1\tf 2}S_{[0,(210)^1]}+{\td 1\tf 10}S_{[1,(300)^1]}+
 {\td 1\tf 4}S_{[1,(210)^1]}+{\td 3\tf 20}S_{[1,(210)^2]}+{\cal I}\nonumber\\[.2cm]
2\Gamma_{K^+ K^-\pi^-\pi^-\pi^+}&=&\Gamma_{K^0\bar K^0\pi^-\pi^-\pi^+}+\Gamma_{K^+ K^-\pi^-\pi^-\pi^+}\nonumber\\
&=& {\td 4\tf 5}S_{[0,(300)^1]}+{\td 1\tf 2}S_{[0,(210)^1]}+{\td 2\tf 5}S_{[1,(300)^1]}+
 {\td 1\tf 4}S_{[1,(210)^1]}+{\td 3\tf 20}S_{[1,(210)^2]}-{\cal I}\nonumber\\[.2cm]
\Gamma_{K^0K^-\pi^+\pi^-\pi^0}&=&S_{[1,(111)^0]}+{\td 1\tf 5}S_{[1,(300)^1]}+ 
{\td 1\tf 2}S_{[1,(210)^1]}+{\td 1\tf 10}S_{[1,(210)^2]}+{\td 2\tf\sqrt{3}}\,{\cal I}\nonumber\\
\Gamma_{K^0K^-\pi^0\pi^0\pi^0}&=&{\td 3\tf 10}S_{[1,(300)^1]}\\
\Gamma_{K^+\bar K^0\pi^-\pi^-\pi^0}&=&{\td 3\tf 5}S_{[1,(210)^2]}.
\end{eqnarray}
Here the partial width $\sum\Gamma_{cc}$ is a function of the interference term:
\begin{equation}
\Gamma_{K\bar K\pi\pi\pi}=\sum_{cc}\Gamma_{cc}=\sum_{sc} S_{sc}+{\td 2\tf\sqrt{3}}\,{\cal I},
\end{equation}
and the  Schwarz's inequality  reads
\begin{equation}\label{schwarz3}
|{\cal I}|\leq{\td 1\tf 2}\sqrt{{\td 3\tf 5}S_{[1,(210)^1]}S_{[1,(210)^2]}}.
\end{equation}\begin{figure}[t]
\begin{center}
\mbox{~}\\[-1cm]
\mbox{\epsfig{file=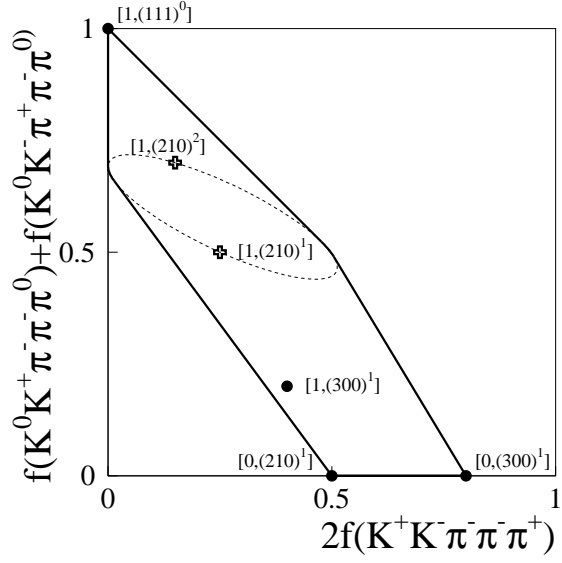,width=8.5cm}}
\end{center}
\caption{Projection of the allowed domain on the plane $x/y$,
$x=\break 2f_{K^+ K^-\pi^+\pi^-\pi^-}$, $y=f_{K^0 K^-\pi^+\pi^-\pi^0}+f_{K^+\bar K^0\pi^-\pi^-\pi^0}$.
The classes of amplitudes are labelled by the isospin values and symmetry classes, 
$[I_{K\bar K},(n_1n_2n_3)^{I_{3\pi}}]$.
}
\end{figure}

The sub-domains are points for the classes $[0,(300)^1]$, $[0,(210)^1]$, $[1,(300)^1]$ and
$[1,(111)^0]$. For the interfering classes  $[1,(210)^1]$  and  $[1,(210)^2]$, the plane of the
two-dimensional sub-domain  
is determined by 
the two relations: $\Gamma_{K^0K^-\pi^0\pi^0\pi^0}=0$ and
%\begin{eqnarray*}
%\lefteqn{
$$2(1+{\td 1\tf\sqrt{3}})\Gamma_{K^+K^-\pi^-\pi^0\pi^0}+2(1-{\td 1\tf\sqrt{3}})\Gamma_{K^+K^-\pi^+\pi^-\pi^-}%}\\
%&&\mbox{~~~~~~~~~~~~~~~~~~~~~~~}
-\Gamma_{K^0K^-\pi^+\pi^-\pi^0}-{\td 1\tf 3}\Gamma_{K^+K^0\pi^-\pi^-\pi^0}=0.$$
%\end{eqnarray*} 
The boundary is given 
by the saturation of the Schwarz's inequality $|{\cal I}|=|{\cal I}|_{max}$. 
The complete, four-dimensional domain is the convex hull of this two-dimensional sub-domain and
the four points. 

Two-dimensional projections can be constructed by the same method that we used in the previous section.
The example shown on Fig. 2 takes $x=2f_{K^+ K^-\pi^+\pi^-\pi^-}$ fraction of decays without 
neutral hadron and $y=f_{K^0K^-\pi^+\pi^-\pi^0}+f_{K^+\bar K^0\pi^-\pi^-\pi^0}$ fraction of decays
with one $\pi^0$ and one $K^0$. The complement $1-x-y$ is the fraction of decays with two or three $\pi^0$'s.
The ellipse bounding  the projection of the two-dimensional sub-domain goes through the two points
$[1,(210)^1]$ and $[1,(210)^2]$. It is tangent to the lines $x=0$ and $x+y=1$, since, due to the interference,
the  contributions of the two classes of amplitudes $[1,(210)^1]$ and  $[1,(210)^2]$ can cancel out in $\Gamma_{K^+K^-\pi^+\pi^-\pi^-}$
or in $\Gamma_{K^+ K^-\pi^0\pi^0\pi^-}$ and there is no contribution to $\Gamma_{K^0K^-\pi^0\pi^0\pi^0}$
from $[1,(210)^1]$ and  $[1,(210)^2]$ amplitudes.
The boundary of the projected domain is made of two arcs of the curve, four tangents and a segment of 
the line $\,y=0$.
Due to the interference, the ratio of the number of decays with two or three $\pi^0$'s over the 
number of decays without $\pi^0$ is only bounded by 0 and 1. The fraction of decays with two or three $\pi^0$'s
is always lower than 1/2.

To distinguish two $\pi^0$ from three $\pi^0$ decays, we can use a third coordinate
$\,z=f_{K^0K^-\pi^0\pi^0\pi^0}$. The three-dimensional domain is the cone having for basis the above 
described contour in the $x/y$ plane and, for vertex, the point $[1,(300)^1]$.
\section{Summary}  
We have presented the complete isospin constraints on the $\tau\to K\bar Kn\pi\nu$ 
decay modes in the space of charge configurations with some details in the cases
$n=2$ and $n=3$.

The geometrical method adopted allows to draw very easily any wanted projection 
of the multi-dimensional domain and hence obtain the most restrictive inequalities 
for a given set of measurements.

\end{document}